\documentclass{article}

\usepackage{arxiv}

\usepackage[utf8]{inputenc} % allow utf-8 input
\usepackage[T1]{fontenc}    % use 8-bit T1 fonts
\usepackage{hyperref}       % hyperlinks
\usepackage{url}            % simple URL typesetting
\usepackage{booktabs}       % professional-quality tables
\usepackage{amsfonts}       % blackboard math symbols
\usepackage{nicefrac}       % compact symbols for 1/2, etc.
\usepackage{microtype}      % microtypography
\usepackage{lipsum}		% Can be removed after putting your text content
\usepackage{graphicx}
\usepackage{doi}

\title{Stochastic Properties of EIP-1559 Basefees}

\author{ \href{https://orcid.org/0000-0000-0000-0000}{\includegraphics[scale=0.06]{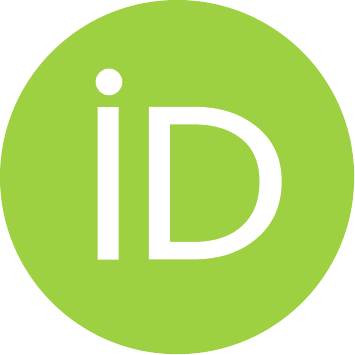}\hspace{1mm}Ian C. Moore, PhD}\\
	Syscoin Researcher\\
	\texttt{imoore@syscoin.org} \\
	\And
	\href{https://orcid.org/0000-0000-0000-0000}{\includegraphics[scale=0.06]{orcid.pdf}\hspace{1mm}Jagdeep Sidhu, MSc} \\
	Syscoin Core Developer\\
	Blockchain Foundry Inc.\\
	\texttt{jsidhu@blockchainfoundry.co} \\
}

\renewcommand{\shorttitle}{EIP-1559 Basefees}

\hypersetup{
pdftitle={Stochastic Properties of EIP-1559 Basefees},
pdfsubject={q-bio.NC, q-bio.QM},
pdfauthor={Ian C. Moore, Jagdeep Sidhu},
pdfkeywords={EIP 1559, Base Fees, Stochastic Processes, Stationarity},
}

\begin{document}
\maketitle

\begin{abstract}
EIP-1559 is a new proposed pricing mechanism for the Ethereum protocol developed to mitigate short-term volatility in demand for transactions. To properly understand this as a stochastic process, it is necessary to develop the mathematical foundations to understand under what conditions the base fee gas price outcomes behave as a stationary process, and when it does not. We believe understanding these mathematical fundamentals is critical to engineering a well-designed system.
\end{abstract}

\keywords{EIP 1559  \and Base Fees  \and Stochastic Processes  \and Stationarity}

\section{Introduction}
The current Ethereum pricing mechanism employs the price auction model where high-value use cases are prioritized over lower ones. The problem with this approach is that as Ethereum has grown more popular, users have difficulty estimating optimal gas fees. In 2018, EIP-1559 was  proposed by Ethereum Founder, Vitalik Buterin, as a major change to Ethereum's transaction fee mechanism, and introduces two different types of fees, which is the base fee and inclusion fee \cite{But19,Rou20}. The idea is to have the base fee relatively stable while the inclusion fee serves as an additional tip offered to compensate miners. This mechanism allows for the base fee to be adjusted after every block in accordance to demand. Hence, when demand for gas is higher than the target gas price, the block size is adjusted upwards, and when demand for gas is lower than the target gas, the block size is adjusted downwards. Blocks are allowed to grow as large as double the target block size.

In this study, we investigate the stochastic properties of EIP-1559 base fees. As the main motivation behind EIP-1559 is to address the volatility and uncertainty associated with gas fees, we look at the mathematical conditions that this mechanism must have to achieve stationarity. We open up this discussion in Section \ref{section:eip_1559} where we go over the EIP-1559 pricing mechanism. Using simulated gas demands, we analyse the stationary properties in Section \ref{section:analyze_stationarity}. In Section \ref{section:gas_demand} we simulate  gas demands.  In Section \ref{section:gas_fees} we analyse stochastic properties of gas fee simulations.  Finally, we close off with our conclusions in Section \ref{section:conclusion}.

\section{EIP-1559 Pricing Mechanism}
\label{section:eip_1559}

The basic premise of the new EIP-1559 proposal begins with setting the base fee which increases when the network capacity exceeds the target per-block gas usage, and decreases when the capacity is below the target. The calculation of the updated basefee, $b_{k+1}$ at the $k+1^{th}$ block, is as follows:

\begin{equation}
b_{k+1} = b_{k}f_{k}
\label{eq:eip1559} 
\end{equation}

\begin{equation}
f_{k} = 1 + \frac{\delta_{k}}{c}
\label{eq:fk} 
\end{equation}
where $\delta_{k} = demand_{k} - Target~Gas~Fee$ and $c = Target ~*~ Base~Fee~Max~Change$; for the purpose of this study, we applied the setup from Table 1, and assumed $\delta_{k}$ to be a normal stationary process. However before we do that, we run the \textit{abm1559} simulator \cite{Mon21} to show that this is the case, under the assumption that new users are Poisson distributed.

\begin{table}[h!]
\centering
\begin{tabular}{ |l|c| } 
\hline
 Constant & Value \\
\hline
Target Gas Fee & 12,500,000 \\
Base Fee Max Change & 50 \\
Initial Basefee & 10,000,000,000\\

\hline
\end{tabular}
\caption{EIP 1559 constants}
\label{table:pow_vs_pos}
\end{table}

\section{Analyse Stationarity}
\label{section:analyze_stationarity}

\textbf{Definition:} A time series $\{x_{t}\}$ is said to be strictly stationary if random vectors $(x_{t_{1}} , ... , x_{t_{n}} )^T $ and $(x_{t_{1}+\tau} , ... , x_{t_{n}+\tau} )^T$ have the same joint distribution for all $\{t \in \mathbb{N}^{0}\}$ and $n,\tau \in \mathbb{N}^{+}$. This is defined as:
\begin{equation}
(x_{t_{1}} , ... , x_{t_{n}})^T \stackrel{d}{=} (x_{t_{1}+\tau} , ... , x_{t_{n}+\tau} )^T,
\label{eq:ar1} 
\end{equation}
where $\stackrel{d}{=}$ means equivalence in distribution.

\subsection{Augmented Dickey Fuller Test}
\label{section:adf}

Given the following higher-order autoregressive processes:

\begin{equation}
\Delta y_{t} = \alpha y_{t-1} + \theta_{1} \Delta y_{t-1}  + ... + \theta_{p} \Delta y_{t-p} + \epsilon_{t},
\label{eq:ar1} 
\end{equation}
the Augmented Dickey Fuller (ADF) test checks the existence of a unit root $\alpha = 1$ (ie, H0: $\alpha = 1$) where the null hypothesis is non-stationary. The unit root in (\ref{eq:ar1}) is characteristic of a time series that makes it non-stationary.

\section{Gas Demand Simulations}
\label{section:gas_demand}

To simulate gas demand $\delta_{k}$ in (\ref{eq:fk}) we use the \textit{abm1559}  simulator \cite{Mon21} where the following assumptions were made: (a) users are Poisson distributed; and (b) every TX consumes 21,000 Gas. For our analysis, we are interested in simulating a year's worth of gas costs, which is infeasible as running the full Ethereum chain simulator consumes a lot of overhead.

Hence, we used the simulator to generate a sample, which was analysed using Exploratory Data Analysis (EDA) techniques, as shown in Figure \ref{fig:eda}. As we can see, the gas demand is behaving as a normal random sample. 

\begin{figure}
\centering
\includegraphics[width=5in]{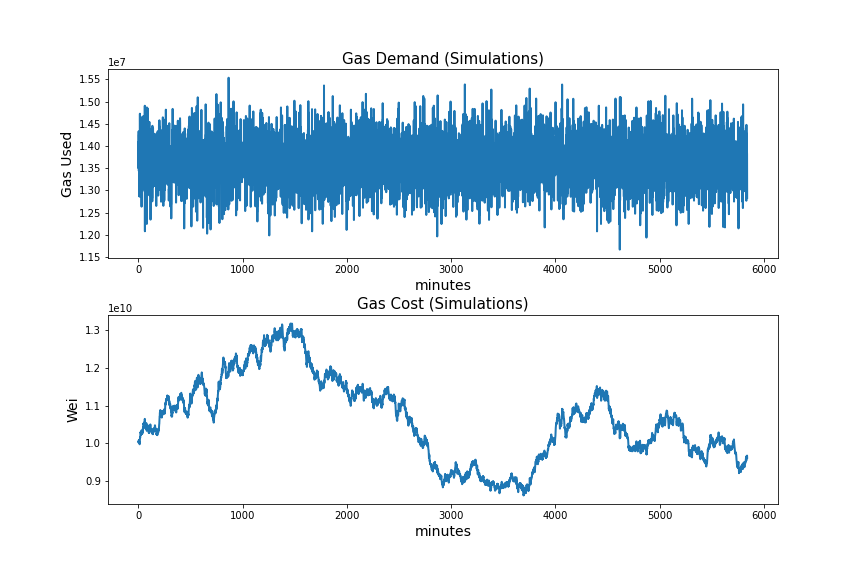}
\caption{Simulations of; (top) gas demands using random samples from a normal distribution; and (bottom) gas costs using (\ref{eq:eip1559}). As we can see, the gas cost simulations (bottom) are behaving like a random walk process} 
\label{fig:gas}
\end{figure} 

\begin{figure}
\centering
\includegraphics[width=5in]{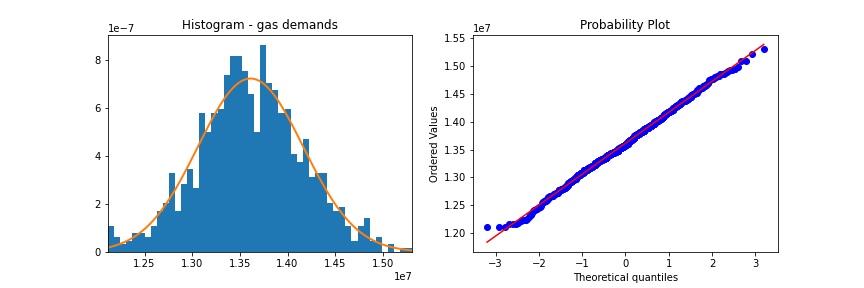}
\caption{Testing normality of gas demands (top image of Fig. \ref{fig:gas}) using; (left) histogram with normal distribution fit of demands; and (right) Normal Q-Q plot indicates sample and theoretical quantiles match} 
\label{fig:eda}
\end{figure} 

We also applied the ADF test to the simulated gas demands, shown in top image of Figure \ref{fig:gas}, and found it to be statistically significant (ie, p-val = 2.33e-29). Hence, rejecting HO, which indicate simulated gas demands to be stationary.

\section{Analyse Gas Fee Simulations}
\label{section:gas_fees}

\begin{figure}
\centering
\includegraphics[width=12 cm]{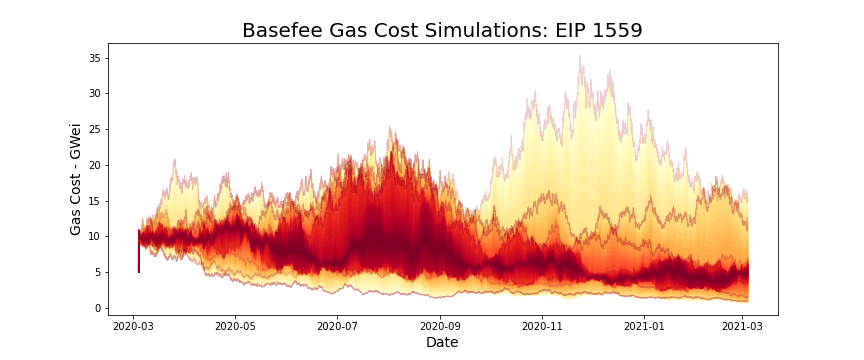}
\caption{Gas cost simulations using EIP-1559; we can visual see the non-stationary behaviour in price over time, which is indicative of a random walk.}  
\label{fig:basefee_simulations}
\end{figure}

In Section \ref{section:gas_demand} we simulated gas demand  $\delta_{k}$ in (\ref{eq:fk}) and found them to be stationary and normally distributed with $\mu$ = 1.36e7, and variance $\sigma^2$ = 5.51e5. To simulate gas prices using EIP-1559 we used these parameter estimates to model a year's worth of gas demands and fed it through the system described in (\ref{eq:eip1559}) and (\ref{eq:fk}). These simulations can be seen in Figure \ref{fig:basefee_simulations}.

We tested gas demands for stationarity. In Table \ref{fig:basefee_simulations} we fail to reject to H0 for all tests indicating insufficient evidence to conclude that the effect of stationarity exists in the gas fee simulations. 
 
\begin{table}
\centering
\begin{tabular}{ |c|c|c|c| } 
\hline
 Simulation & ADF Statistic & p-value & Outcome \\
\hline
1 & -1.506 & 0.531 & not stationary \\
2 & -1.260 & 0.647 & not stationary \\
3 & -2.170 & 0.217 & not stationary \\
4 & -2.432 & 0.133 & not stationary \\
5 & -1.844 & 0.359 & not stationary \\
6 & -1.323 & 0.618 & not stationary \\
7 & -1.645 & 0.459 & not stationary \\
8 & -1.662 & 0.451 & not stationary \\
9 & -1.124 & 0.705 & not stationary \\
10 & -1.930 & 0.318 & not stationary \\
\hline
\end{tabular}
\caption{ADF test on basefee simulations from Figure \ref{fig:basefee_simulations} to test null hypothesis (H0) that a unit root is present; when unit root is present, then sample is considered to be non-stationary. As we can see, we fail to reject H0 for all ten tests indicating insufficient evidence to conclude that the effect of stationarity exists.}
\label{table:pow_vs_pos}
\end{table}

\subsection{Problem Statement}
\label{section:problem_statement}
For sake of comparison, let's look at the stationarity for a standard first-order auto-regressive process  (ie, AR(1)) defined as:

\begin{equation}
x_{t} = \mu + \alpha (x_{t-1} -\mu) + \epsilon_{t-1} ~~~~~~~~ \epsilon_{t} \sim IID~N(0,\sigma^2)
\label{eq:ar1} 
\end{equation}
The above process is stationary when $|\alpha|$<1, hence has a single root 1/$\alpha$ outside the unit circle. The asymptotic stationary distribution can be shown to be:
\begin{equation}
x_{\infty} \sim N(\mu,\frac{\sigma^2}{1-\alpha^2}),
\label{eq:x_infinity} 
\end{equation}
when $\sigma$>0. Considering the (state space like) simularities to the AR(1) process to the system of the system of (\ref{eq:eip1559}) and (\ref{eq:fk}); the question is what are the constraints (if any) that time varying process $\delta_{t}$ must meet to ensure stationary outcomes for the basefees.

\subsection{Random Coefficient Autoregressive Models}
\label{section:rca}

Considering the problem statement in the previous section, we look to another relative to the AR(1) process, namely the Random Coefficient Autoregressive of order 1, or RCA(1) process. The RCA(1) is similar to the AR(1) process in the sense that the parameter $\alpha$ is allowed to vary with time. This is quite similar to the EIP-1559 system of (\ref{eq:eip1559}) and (\ref{eq:fk}). The RCA(1) model is given by:

\begin{equation}
x_{t} = \alpha + \beta_{t}x_{t-1} + \epsilon_{t},
\label{eq:rca1}
\end{equation}
where ${\epsilon_{t}}$ is an independent sequence of random variables with 0 mean and variance $\sigma^2 > 0$; ${\beta_{t}}$ is an independent sequence of random variables with mean $\mu_{\beta}$ and variance $\sigma_{\beta}^2$ of the random coefficient $\beta_{t}$, and variance $\sigma^2$ of the error $\epsilon_{t}$. Given (3) and (4) it is obvious that when $\sigma_{\beta}^2 = 0$, the RCA(1) in (5) becomes a AR series, and becomes a random walk when $\mu_{\beta} = 1$ and $\sigma_{\beta}^2 = 0$, hence non-stationary. 

\begin{figure}
\centering
\includegraphics[width=5in]{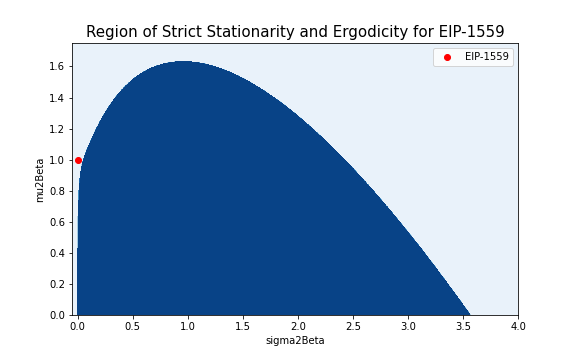}
\caption{Region of strict stationarity and ergodicity for RCA(1) determined via (\ref{eq:thm}). The annotated red dot represents where our setup (Table 1) is in relation to the region where an RCA(1) process behaves as a strictly stationary process. Based on this we can see that our setup was non-stationary.} 
\label{fig:strict_stationarity}
\end{figure}

EIP-1559 (1) fits into the framework of a Random Coefficient Autoregressive (RCA) model when $\epsilon$ is negligibly small. This is good news, as it allows us to apply the body of work on RCA processes to this problem. The next question is; under what conditions does an RCA(1) process exhibit stationary behaviour. If we know this, then we can apply this understanding to the EIP-1559 mechanism. To help with our understanding regarding this, we apply Wang's Theorem \cite{Wan03}; see Figure \ref{fig:strict_stationarity}.

\textbf{Theorem} Consider RCA(1) model (5) with ${\beta_{k},\epsilon_{k}}$, which is identically normally distributed.  Then the sufficient condition for the existence of a strictly stationary and ergodic solution is that:

\begin{equation}
\ln(\sigma_{\beta}^2) < \varsigma + \ln(2) - 2 \int_{0}^{1}\frac{1 - \exp[-\lambda(1-w^2)]}{1-w^2}dw,
\label{eq:thm} 
\end{equation}
where $\varsigma$ $\approx$ 0.57721, denotes Euler’s constant and $\lambda$ = $\mu_{\beta}^2$/ $2\sigma_{\beta}^2$

\textbf{Proof:} See Theorem 2 in \cite{Wan03}.

\section{Conclusion}
\label{section:conclusion}

The purpose of this study was to understand under what conditions the EIP-1559 pricing mechanism would behave as a stationary process. We believe this as a critical first step to understanding the constraints to engineering a well-designed system. Given the enormity of the Ethereum project, understanding these fundamentals are needed for setting up well-designed simulation experiments to help capture all the edge cases that may arise so that the number of updates and patches are minimized prior to release.

It was determined that under the parameter setting used in this study, that the outcomes behaved as a random walk, which is a non-stationary process. The RCA(1) was used as the mathematical framework to help understand this, as we can go to the literature to help guide our understanding of this kind of process. This conclusion was based on a series of simulations which we tested using the ADF, and Wang's Theorem indicating the region of stationarity for an RCA(1) process. 

Ideally, we would like EIP-1559 to behave as a stationary process. Hence, more investigation is required to understand the setup that is necessary to achieve these conditions before proper simulation studies are implemented.

\end{document}